\documentclass[preprintnumbers,aps,pre,twocolumn,superscriptaddress,showpacs,amsmath,amssymb]{revtex4}

\usepackage{color}
\usepackage{graphicx}
\usepackage{dcolumn}
\usepackage{bm}

\newcommand{\felastic}{f_{\mathrm{elastic}}}

\newcommand{\NW}{_{\mathrm{NW}}}

\newcommand{\nvec}{\mathbf{n}}

\newcommand{\Qvec}{\mathbf{Q}}
\newcommand{\Qij}{Q_{ij}}

\newcommand{\Fcal}{\mathcal{F}}

\newcommand{\dd}{\mathrm{d}}

\DeclareMathOperator{\Tr}{Tr}

\begin{document}


\title{Generalized Berreman's model of the elastic surface free energy 
of a nematic liquid crystal on a sawtoothed substrate}

\author{O. A. Rojas-G\'omez}
\affiliation{Departamento de F\'{\i}sica At\'omica, Molecular y Nuclear, Area de
F\'\i sica Te\'orica
Universidad de Sevilla,
Apartado de Correos 1065, 41080 Sevilla, Spain
}%
\author{J. M. Romero-Enrique}
\affiliation{Departamento de F\'{\i}sica At\'omica, Molecular y Nuclear, Area de F\'\i sica Te\'orica
Universidad de Sevilla, 
Apartado de Correos 1065, 41080 Sevilla, Spain
}%

\date{\today}

\begin{abstract}

In this paper we present a generalization of Berreman's model for 
the elastic contribution to the surface free-energy density of a nematic 
liquid crystal in presence of a sawtooth substrate which favours homeotropic 
anchoring, as a function of the wavenumber of the surface
structure $q$, the tilt angle $\alpha$ and the surface anchoring strength
$w$. In addition to the previously reported non-analytic contribution 
proportional to $-q\ln q$, due to the nucleation of disclination lines at the
wedge bottoms and apexes of the substrate, the next-to-leading contribution
is proportional to $q$ for a given substrate roughness, in agreement with 
Berreman's predictions. We characterise this term, finding that it has two
contributions: the deviations of the nematic director field with respect to
a reference field corresponding to the isolated disclination lines, and 
their associated
core free energies. Comparison with the results obtained from the Landau-de 
Gennes model shows that our model is quite accurate in the limit $wL>1$, 
when strong anchoring conditions are effectively achieved. 

\end{abstract}

\maketitle

\section{Introduction}

The behaviour of nematic liquid crystals in the presence of microstructured
substrates has been subject of intensive research in the recent 
times \cite{lee,kim,ferjani}. 
This problem has practical applications such as the design of 
zenithally bistable devices \cite{Brown_2000,Parry-Jones1,Parry-Jones2,
davidson,evans}, or the trapping of colloidal particles in specified sites 
\cite{nuno,ohzono}. It is well known that the nematic director field, in
presence of the structured substrate, may be distorted, leading to an 
elastic contribution to the free energy. Since the seminal work by Berreman
\cite{berreman,gennes}, this problem has been extensively studied and
generalized in the literature \cite{barbero1,barbero2,barbero3,Brown_2000,
Kitson_2002,Fukuda,patricio5,harnau,kondrat,harnau2,harnau3,barbero4,Yi_2009,
poniewierski,romero}. Wetting and filling transitions by nematic on these
grooved surfaces have also been studied \cite{bramble,patricio,patricio2,
patricio4}. When the substrate presents cusps, topological defect
nucleate on them \cite{barbero1,barbero2,barbero3,poniewierski,romero}, and
Berreman's expression of the elastic contribution to the free-energy density,
which is proportional to the wavenumber of the substrate structure $q$, 
breaks down because of the emergence of a non-analytical contribution 
proportional to $-q\ln q$ associated to the nucleated defects \cite{romero}.
This result constrasts with the phenomenology observed in smooth substrates,
as sinuosidal substrates,
where in absence of topological defects a suitable generalization of 
Berreman's model works \cite{barbero4,patricio4}. 

In this paper we will complete the characterization of the elastic contribution
to the surface free-energy density for sawtoothed substrates \cite{romero}.
Beyond the $-q\ln q$ term previously mentioned, we find that the 
next-to-leading contribution follows Berreman's scaling with the wavenumber
$q$. The origin of this term is twofold: the deviations of the nematic
director field with respect to the distortions imposed by the presence of the 
nucleated disclination lines, and the defect core contributions. We estimate
both contributions, finding a fairly good agreement with the reported values
in Ref. \cite{romero}.

The paper is organized as follows. The problem is set up in Section II, where
we identify the different contributions to the elastic contribution to the
surface free-energy density. Sections III and IV are devoted to the estimation
of these contributions, and the obtained results will be discussed in
Section IV. We will end up with the conclusions in Section V.

\section{The model}

\begin{figure}
\centerline{\includegraphics[width=.95\columnwidth]{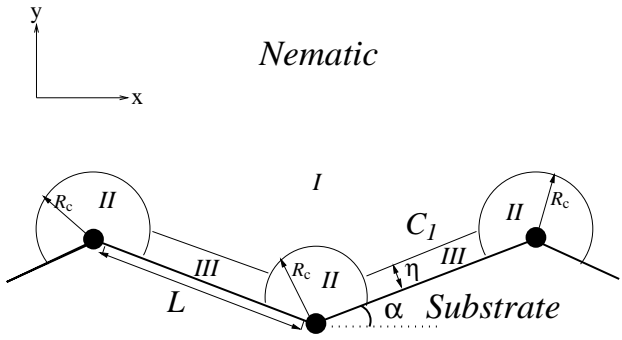}}
\caption
{Schematic picture of the geometry of the system, characterized by the
side length $L$ and the angle $\alpha$. The different regions $I$, $II$
and $III$ are outlined. See text for explanation.}
\label{fig1}
\end{figure}

We consider a nematic liquid  
in contact with a sawtooth substrate characterized by the 
angle $\alpha$ and the length side $L$ (see Fig. \ref{fig1}). 
The substrate favours homeotropic anchoring of the molecules. 
We assume traslational symmetry along $z$ axis and a periodic distribution 
of wedges and cusps along the $x$ axis. Under these conditions, the nematic 
director field $\nvec(\mathbf{r})$ shows only in-plane distortions 
\cite{romero}, so it can be parametrized by using the angle $\theta$ between 
the local director and the $y$ axis, yielding 
$\nvec(\mathbf{r})=(-\sin\theta(\mathbf{r}),\cos\theta(\mathbf{r}),0)$. 
Far from the substrate, the bulk nematic phase orients homogeneously along 
either the $x$ axis (perpendicular texture $N^\perp$) or the $y$ axis 
(parallel texture $N^\parallel$). 
The nematic order of the system can be locally represented by a traceless 
symmetric order parameter second-rank tensor $\Qvec$, with Cartesian components
$\Qij = \frac 3 2 S[n_in_j - \frac 1 3 
\delta_{ij}]+ \frac 1 2 B [l_il_j - m_i m_j]$, where $S$ is the nematic 
order parameter, which measures the orientational ordering along the
nematic director, and $B$ the biaxiality parameter, which
measures the ordering of the molecules on the orientations
perpendicular to $\mathbf{n}$, characterized by the eigenvectors 
$\mathbf{l}$ and $\mathbf{m}$.

As in previous works \cite{romero,patricio,patricio2}, the system will be 
described within 
the Landau-de Gennes (LdG) framework. The order parameter tensor profile 
is obtained by minimizing the surface free-energy density functional 
\begin{eqnarray}
&& {\cal F}= 
\frac{1}{\lambda}\int_{-\lambda/2}^{\lambda/2} \dd x 
\int_{|x|\tan\alpha}^\infty \dd y 
\Bigg[\frac{3-2\tau}{4\tau-3}\left(\frac{2\Tr  \tilde\Qvec^2}{3}-1
\right) \nonumber\\&& - \frac{2}{4\tau-3}\left(\frac{4}{3} \Tr
\tilde\Qvec^3 -1\right)+ \frac{\tau}{4\tau-3}\left(\frac{4}{9} 
[\Tr\tilde\Qvec^2]^2 -1\right) \nonumber\\
&&+ \frac {1}
{3+2\kappa}[ \partial_k \tilde Q_{ij} \partial_k \tilde Q_{ij} + 
\kappa  \partial_j 
\tilde Q_{ij}
\partial_k \tilde Q_{ik}] \Bigg]\nonumber \\ && 
-\frac{1}{\lambda}\int_{-\lambda/2}^{\lambda/2}\dd x \frac 2 3 w \Tr
[\tilde\Qvec\cdot\tilde\Qvec_{\mathrm{surf}}]
\Big |_{y=|x|\tan\alpha}
\label{free_energy}
\end{eqnarray}
where $\lambda\equiv 2L\cos\alpha$ is the substrate periodicity wavelength,
$\tilde\Qvec=\Qvec/S_b$, where $S_b=S_b(T)$ is the bulk value of the nematic
order parameter at the temperature $T$, and $\tau=S_b(T)/S_b(T_{NI})$ is
the ratio between the nematic order parameter at the temperature $T$ and
the nematic-isotropic transition temperature $T_{NI}$.
The positions are measured in units of $\sqrt{2}\xi$, where $\xi$ is the 
nematic correlation length along the local nematic director axis. Finally
$\kappa$ is the ratio between the relevant elastic parameters ($\kappa>-3/2$) 
and $w$ the (reduced) anchoring strength. Homeotropic alignment of the nematic 
is favoured by setting $\tilde\Qvec_{\mathrm{surf}}=(3\boldsymbol{\nu}\otimes
\boldsymbol{\nu}-1)/2$, being 
$\boldsymbol{\nu}=((x/|x|)\sin\alpha,-\cos\alpha,0)$ the outwards normal 
vector to the substrate. The global minimum of the functional 
Eq. (\ref{free_energy}) yields the mean-field equilibrium surface 
free-energy density, $f$, in appropiated units. 
The contribution due to the elastic 
deformations induced by the substrate structure, $\felastic$, can be obtained 
as $\felastic=f-r\sigma_{NW}(w)$, where the surface roughness is 
$r=1/\cos\alpha$ and $\sigma_{NW}(w)$ is the nematic-flat substrate
interfacial tension. In a similar way as it was obtained at nematic isotropic
coexistence in Ref. \cite{patricio4}, the interfacial tension for the 
LdG model has the expression:
\begin{eqnarray}
&&\sigma\NW=
\frac{\sqrt{2}\sqrt{(\tau-1)(\tilde{S}(0)+1)^2+\tilde{S}(0)^2}}
{6\tau^2\sqrt{4\tau-3}}
\nonumber\\ &&\times
\left(2\tau^2\tilde{S}(0)^2-\tau(2\tau+1)\tilde{S}(0)-4\tau^2+7\tau-3
\right)\nonumber\\
&&-\frac{\sqrt{2}\sqrt{4\tau-3}(-4\tau^2+6\tau-3)}{6\tau^2\sqrt{4\tau-3}}
\nonumber\\
&&-\sqrt{2}\frac{1-3\tau+2\tau^2}{2\tau^{5/2}\sqrt{4\tau-3}}\nonumber\\
&&\times\ln\left(\frac{\tau\tilde{S}(0)+
\tau-1+\sqrt{\tau}\sqrt{(\tau-1)(\tilde{S}(0)+1)^2+\tilde{S}(0)^2}
}{2\tau-1+\sqrt{\tau}
\sqrt{4\tau-3}}\right)\nonumber\\
&&-w\tilde{S}(0)
\label{sigmanw}
\end{eqnarray}
where $\tilde{S}(0)=S(0)/S_b$ and $S(0)$ is the nematic order parameter
at the substrate. The value of $\tilde{S}(0)$ is obtained as the largest 
solution of the equation:
\begin{equation}
(3-2\tau)\tilde{S}^2(0)-2\tilde{S}^3(0)+\tau\tilde{S}^4(0)=1-\tau+
\frac{4\tau-3}{2}w^2
\label{s0}
\end{equation}
At coexistence $\tau=1$, and Eq. (\ref{sigmanw}) reduces to:
\begin{equation}
\sigma\NW=\frac
{\sqrt{2}(2\tilde{S}(0)+1)(\tilde{S}(0)-1)^2}{6}-w\tilde{S}(0)
\label{sigmanwcoex}
\end{equation}
where $\tilde{S}(0)=(1+\sqrt{1+2\sqrt{2}w})/2$. 
 
A systematic study of this system via full minimization of the LdG functional
was done in Ref. \cite{romero}. 
For this purporse, the functional Eq. (\ref{free_energy}) was numerically
minimized by using a conjugate-gradient method. The numerical discretization 
of the continuum problem is performed with a
finite element method \cite{zienkiewicz} 
combined with adaptive meshing in order to resolve the
different length scales that may emerge in the problem
\cite{patricio3}. It was found that the $N^\perp$
texture has lower free energy if $\alpha<\pi/4$ owing to lesser distortion.
Conversely, the $N^\parallel$ texture has lower free energy for 
$\alpha>\pi/4$, in agreement with earlier predictions \cite{barbero1,
barbero2,barbero3,poniewierski}. For large $wL$, strong anchoring conditions
are effectively achieved, leading to the nucleation of  
disclination lines characterized by non-half-integer winding numbers 
along the ridges and wedges of the substrate \cite{barbero1,
barbero2,barbero3,poniewierski,romero}. As a consequence, the elastic 
contribution to the surface free-energy density has the following scaling
\cite{romero}
\begin{equation}
\felastic\approx
- \frac{\mathcal{K}(\alpha)}{2\pi}q\ln \frac{q\cos\alpha}{\pi} + 
\frac{q}{2\pi}B(\alpha,w)
\label{scaling}
\end{equation}
where $q=2\pi/\lambda=\pi/L\cos\alpha$ is the substrate periodicity wavenumber,
$\mathcal{K}(\alpha)$ is defined as:
\begin{equation}
\mathcal{K}(\alpha)=\begin{cases}
\frac{K\pi \alpha^2}{\left(\frac{\pi}{2}
\right)^2-\alpha^2} & N^\perp\ \textrm{texture}\\
\\
K\pi\frac{\frac{\pi}{2}-\alpha}{\frac{\pi}{2}
+\alpha} & N^\parallel\ \textrm{texture}\\
\end{cases}
\label{defkalpha}
\end{equation}
being $K=(9/2)(2+\kappa)/(3+2\kappa)$ the reduced bulk elastic constant 
associated to bend and splay distortions. From the numerical results, 
the function $B$ 
is found to depend on the substrate rougness (i.e. $\alpha$) and nematic 
texture, as well as the anchoring $w$, but asymptotically \emph{not} on $L$ 
for large $wL$. However, as $L$ increases, the complete minimization becomes
very time-consuming. On the other hand, we do not get information about the 
origin of $B(\alpha,w)$. 

In this paper we will introduce an alternative way to obtain $\felastic$ from
the functional Eq. (\ref{free_energy}). We divide the minimization domain into
three regions (see Fig. \ref{fig1}): most of the domain will correspond 
to the region $I$, formed
by the points which are far enough from the substrate. The neighbourhood of
the substrate will be split into two regions: the region $II$, formed by the
union of the circular sections of radii $\xi<R_c\ll L$ centered at each wedge
and apex; and region $III$, which are the points which are at a distance 
smaller than $\eta \sim \xi$ to the substrate, but at distances larger than
$R_c$ from any substrate ridge. Our hypothesis is that, for large $wL$, the
minimization of the surface free-energy functional restricted to each region
(subject to appropriated boundary conditions), gives a good approximation
to the complete minimization of $\Fcal$. On the other hand, we anticipate that
this analysis will give us some insight in the different contributions
to $B(\alpha,w)$. 

We start with the minimization of 
region $III$. As it was argued in Ref. \cite{romero}, large $wL$ leads to 
strong anchoring conditions. So, in order to minimize the surface free-energy
density, we impose to the angle field $\theta(\mathbf{r})$ to be constant
along its boundary, so the nematic director field is homogeneous and equal to
the normal to the substrate. Consequently, the minimization of the free-energy
functional in this region will lead to a homogeneous director field normal
to the substrate, although the nematic order 
parameter $S$ at each point will depend on its distance to the substrate. We 
impose the following fixed boundary conditions for $S$: $S=S_b$ at the 
boundary between regions $I$ and $III$, and the equilibrium nematic order 
parameter profile for the flat wall case at the boundary between regions
$II$ and $III$. Assuming that $\eta$ is large enough, this
situation is completely equivalent to the flat case, so the minimum value
of the surface free-energy density in this region, $f_{III}$, will 
be $\sigma_{NW}(1-2R_c/L)/\cos\alpha$, with corrections of order of
$\exp(-\eta/\xi)$. Next Sections will be devoted to the
evaluation of the minimum values of the surface free-energy functionals at
the remaining regions, $f_I$ and $f_{II}$.

\section{Evaluation of $f_I$}

The variations of $S$ are restricted to the neighborhood of the 
substrate of a width typically of order of $\xi$, and inside the defect 
cores. So, in region $I$, $S$ takes the bulk value $S_b$ everywhere 
\cite{romero}. Thus, the surface free-energy functional to minimize in region
$I$ reduces to a Frank-Oseen functional: 
\begin{equation}
\mathcal{F}_I\approx 
\frac{K}{2\lambda}\int_{I} \dd x \dd y 
|\boldsymbol \nabla \theta|^2 
\label{mfofreeenergy}
\end{equation}
where the integration is restricted to region $I$, $K$ is the reduced 
elastic constant, and $\theta$ is the orientation field.
The minimization of Eq. (\ref{mfofreeenergy}) yields to the Laplace equation
for $\theta$, $\nabla^2\theta(\mathbf{r})=0$. In the far field, we impose
Dirichlet boundary conditions $\lim_{y\to \infty} \theta(\mathbf{r})=
\alpha_\infty$, where $\alpha_\infty=0$ for the $N^\perp$ texture and
$\alpha_\infty=\pi/2$ for the $N^\parallel$ texture. Along the contours $x=
\pm \lambda/2$, we should impose periodic boundary conditions. However, 
we impose instead Dirichlet boundary conditions $\theta=\alpha_\infty$ along 
these contours, as we know from the full LdG model minimization that these
are the conditions satisfied by the mean-field solution \cite{romero}. Finally, 
we assume strong anchoring conditions along the boundary between regions 
$I$ and $III$: $\theta(x,y=x^2\tan\alpha/|x|,z)=\alpha_\infty+(x/|x|)
(\alpha-\alpha_\infty)$. As discussed 
above, this condition will be accurate if $wL\gg 1$.  
Their contribution to the free-energy density, $f_I$, 
comes from a contour integration of the mean-field solution via \cite{romero}:
\begin{eqnarray}
f_I &=&
\frac{K(\alpha-\alpha_\infty)}{\lambda} \int_{C_1} \boldsymbol{\nu} \cdot
\boldsymbol{\nabla} \theta \dd \mathbf{s}
\label{contour_integral}
\end{eqnarray}
where $C_1$ is the contour parallel to the boundary between regions $I$ and
$III$ between a wedge and apex (see Fig. \ref{fig1}).

As argued in Refs. \cite{barbero1,barbero2,barbero3,romero}, 
the presence of cusps in the substrate 
induces the formation of disclination lines, which are the responsible of the 
non-Berreman scaling of the elastic contribution to the surface free-energy 
density. In this Section, we will complete that analysis, evaluating the
next-to-leading contribution. 

\subsection{Singular contribution}

The solution $\theta(\mathbf{r})$ to the Laplace equation subject to the 
boundary conditions mentioned above can be split into two terms: a singular 
contribution $\theta_s(\mathbf{r})$ due to the periodic array of disclination
lines nucleated at the ridges of the substrate, and a non-singular contribution
$\theta_{ns}(\mathbf{r})$. A representation of the singular contribution for
each texture is given by \cite{romero}:
\begin{eqnarray}
\theta_s^\perp&=&\frac{-\alpha}{\frac{\pi}{2}-\alpha}\Bigg(-\arctan
\left[\tan\frac{qx}{2}\coth\frac{qy}{2}\right]\nonumber\\
&+&\arctan
\left[\tan\frac{qx}{2}\right]\Bigg)
\label{thetasingperp}\\
&+&\frac{\alpha}{\frac{\pi}{2}+\alpha}\Bigg(-\arctan
\left[\tan\frac{qx}{2}
\tanh\frac{q(y-L\sin\alpha)}{2}\right]
\nonumber\\
&+&\arctan\left[\tan\frac{qx}{2}\right]\Bigg)
\nonumber\\
\theta_s^\parallel&=& \frac{\pi}{2}+
\Bigg(-\arctan
\left[\tan\frac{qx}{2}\coth\frac{qy}{2}\right]\nonumber\\
&+&\arctan
\left[\tan\frac{qx}{2}\right]\Bigg)
\label{thetasingpar}\\
&-&\frac{\frac{\pi}{2}-\alpha}{\frac{\pi}{2}+\alpha}\Bigg(-\arctan
\left[\tan\frac{qx}{2}
\tanh\frac{q(y-L\sin\alpha)}{2}\right]
\nonumber\\
&+&\arctan\left[\tan\frac{qx}{2}\right]\Bigg)
\nonumber
\end{eqnarray}
Their contribution to the free-energy density, $f_I^s$, comes from a contour 
integration of these solutions:
\begin{eqnarray}
f_I^s &=&
\frac{K(\alpha-\alpha_\infty)}{\lambda} \int_{C_1} \boldsymbol{\nu} \cdot
\boldsymbol{\nabla} \theta_{s} \dd \mathbf{s}
\label{contour_integral2}
\end{eqnarray}
In Ref. \cite{romero} it was estimated the large-$L$ behaviour of Eq.
(\ref{contour_integral2}), leading to the non-Berreman term ${\mathcal{K}}
(\alpha)q\ln(L/R_c)$ term. However, after some algebra it is possible to
obtain explicitely $f_I^s$ from the solutions Eqs. (\ref{thetasingperp}) 
and (\ref{thetasingpar}) via Eq.(\ref{contour_integral2}) as:
\begin{equation}
f_I^s=\frac{\mathcal{K}(\alpha)q}{2\pi} 
\left(-\ln qR_c +\ln\left[2\cosh\left(\frac{\pi}
{2}\tan\alpha\right)\right]-\alpha\tan\alpha \right)
\label{fIsing}
\end{equation}
where ${\mathcal{K}}(\alpha)$ depends on the texture and substrate
geometry as Eq. (\ref{defkalpha}) and we neglected terms of order 
$q^3 (R_c)^2$ and $q\eta$. 
As the dependence on the nematic texture comes from 
$\mathcal{K}(\alpha)$, 
$f_I^s$ will be minimum for the $N^\perp$ (resp. $N^\parallel$)
texture for $\alpha<\pi/4$ (resp. $\alpha>\pi/4$).

Two remarks are pertinent at this point. First, we note that, although $f_I^s$
may depend on the explicit representation of the singular solution, the leading
non-Berreman contribution is independent on this representation.  
The reason for this fact is that this leading contribution arises from the
behaviour close to the wedges and apexes of $\theta_s$, which must 
asymptotically approach to the corresponding to an isolated disclination 
line \cite{romero}. Secondly, the next-to-leading contribution 
gives a first contribution to $B(\alpha,w)$, which we will denote as
$B_{I,s}(\alpha)$, given by the expression:
\begin{equation}
B_{I,s}(\alpha)={\mathcal{K}}(\alpha)\left(\ln\left[\frac{2}{\pi}
\cosh\left(\frac{\pi}
{2}\tan\alpha\right)\cos\alpha\right]-\alpha\tan\alpha\right) 
\label{defbIs}
\end{equation}

\subsection{Non-singular contribution}

The non-singular part of the mean-field solution, $\theta_{ns}$, is solution
of the Laplace equation $\nabla^2 \theta_{ns}=0$, subject to the boundary 
conditions $\theta_{ns}=0$ in the far field, i.e. $y\to \infty$ and 
along the boundaries $x=\pm \lambda/2$. On the other hand, $\theta_{ns}=
\alpha-\theta_s$ along $C_1$. We do not have an explicit expression for
$\theta_{ns}$ (however, there is an implicit expression via a 
Schwarz-Christoffel transformation \cite{barbero1}, see below), so we
have to resort to numerical methods. We have used two different techniques: 
a finite element method, analogous to the method outlined in the Section II
to solve the LdG model, but minimizing instead the functional 
Eq. (\ref{mfofreeenergy}) subject to the boundary conditions for 
$\theta_{ns}$ mentioned above; and as an alternative, the boundary 
element method \cite{brebbia,katsikadelis}. In this technique the solution
$\theta_{ns}$ inside the region $I$ can be written as:
\begin{eqnarray} 
&&\theta_{ns}(\mathbf{r})=\oint_{\partial I} \dd \mathbf{s} \Bigg( 
[\boldsymbol{\nu}(\mathbf{s})\cdot \boldsymbol{\nabla}_{\mathbf{s}} 
\theta_{ns}(\mathbf{s})]
G(\mathbf{s},\mathbf{r})\nonumber\\
&&-\theta_{ns}(\mathbf{s})
\left[\boldsymbol{\nu}(\mathbf{s})\cdot \boldsymbol{\nabla}_{\mathbf{s}}
G(\mathbf{s},\mathbf{r})\right]\Bigg)
\label{bem1}
\end{eqnarray}
where the contour integral over the boundary $\partial I$ of region $I$ is
counter-clockwise, $\boldsymbol{\nu}(\mathbf{s})$ is the outwards normal
to the boundary at $\mathbf{s}$ and $G(\mathbf{s},\mathbf{r})$ is the
fundamental solution of the Laplace equation 
$G(\mathbf{s},\mathbf{r})=-\ln\left(|\mathbf{s}-\mathbf{r}|\right)/2\pi$.
As we impose Dirichlet boundary conditions, the second term in the right-hand
side of Eq. (\ref{bem1}) is known. On the other hand, the normal derivative of 
$\theta_{ns}$ at the boundary is obtained by solving the integral 
equation \cite{brebbia,katsikadelis}:
\begin{eqnarray} 
&&\oint_{\partial I} \dd \mathbf{s} 
[\boldsymbol{\nu}(\mathbf{s})\cdot \boldsymbol{\nabla}_{\mathbf{s}} 
\theta_{ns}(\mathbf{s})]
G(\mathbf{s},\mathbf{r})=\frac{\theta_{ns}(\mathbf{r})}{2}\nonumber\\
&&+\oint_{\partial I} \dd \mathbf{s} \theta_{ns}(\mathbf{s})
\left[\boldsymbol{\nu}(\mathbf{s})\cdot \boldsymbol{\nabla}_{\mathbf{s}}
G(\mathbf{s},\mathbf{r})\right]
\label{bem2}
\end{eqnarray}
where now $\mathbf{r}\in \partial I$. In order to solve Eq. (\ref{bem2}), we
discretize the boundary in a set of straight segments (the boundary elements). 
We use the constant boundary element approach \cite{katsikadelis}, so we 
assume that both $\theta_{ns}$ and its normal derivative are constants along
each boundary element. Introducing this approximation to Eq. (\ref{bem2}),
we obtain a set of linear algebraic equations for the normal derivatives
of $\theta_{ns}$. Once we solve this set of equations, and introducing the 
same approximation in Eq. (\ref{bem1}) we obtain the non-singular orientational
field $\theta_{ns}$ inside region $I$.   

Once evaluated $\theta_{ns}$, its contribution to the surface free-energy 
density can be obtained from:
\begin{eqnarray}
&& f_I^{ns} =
\frac{K(\alpha-\alpha_\infty)}{\lambda} \int_{C_1} \boldsymbol{\nu} \cdot
\boldsymbol{\nabla} \theta_{ns} \dd \mathbf{s} 
\label{nsfreeenergy1}\\
&& =\frac{K}{2\lambda}
\int_I \dd \mathbf{r} |\boldsymbol{\nabla} \theta_{ns}|^2
+\frac{K}{\lambda}
\int_{C_1} (\alpha-\theta_s(\mathbf{s}))\boldsymbol{\nu} \cdot
\boldsymbol{\nabla} \theta_{s} \dd \mathbf{s} 
\label{nsfreeenergy2}
\end{eqnarray}
where the first result is more appropriated for the boundary element technique,
while the second is more appropriate for the finite element method (note
that the last term in the second result can be evaluated numerically with
high accuracy by standard methods as we
know analytically $\theta_s(\mathbf{r})$). 

\begin{figure}
\centerline{\includegraphics[width=0.5\textwidth]{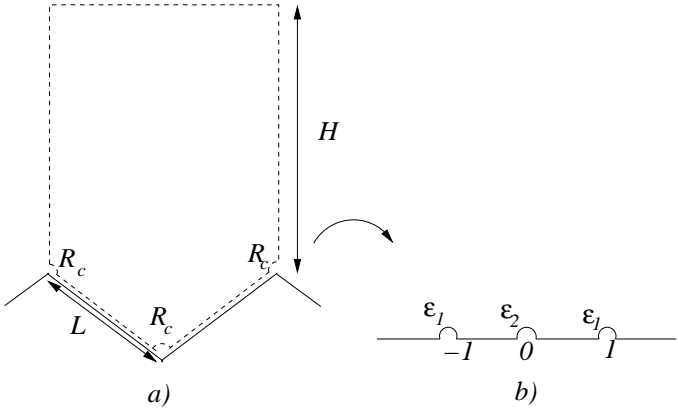}}
\caption
{Left panel: Minimization cell for the evaluation of $f_{I,ns}$. Right panel:
Mapping of the minimization cell under the Schwarz-Christoffel transformation
in the limit $H\to \infty$.}
\label{fig2}
\end{figure}

\begin{figure}
\centerline{\includegraphics[width=.95\columnwidth]{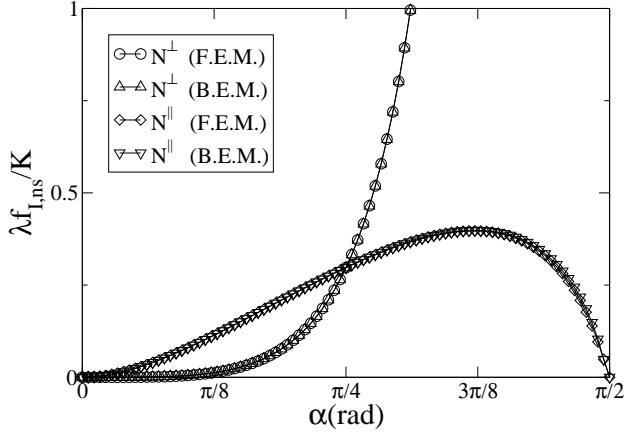}}
\caption
{Plot of $\lambda f_{I,ns}/K$ as a function of $\alpha$ for the $N^\perp$ 
and $N^\parallel$ textures, by using the finite-element method (F.E.M.) 
and the boundary element method (B. E. M.).
}
\label{fig3}
\end{figure}

\begin{figure}
\centerline{\includegraphics[width=.95\columnwidth]{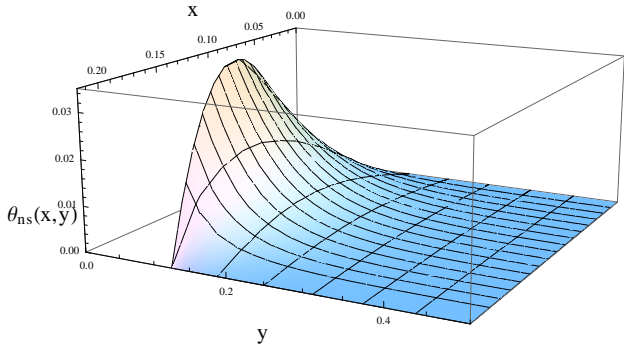}}
\caption
{(Color online) 
Plot of $\theta_{ns}$ as a function of $x$ and $y$ obtained from the
finite-element method for $\alpha=\pi/6$ and $L=0.25$. For the sake
of clarity it is only represented in half a minimization cell 
$0<x<L\cos\alpha$. 
}
\label{fig4}
\end{figure}

The numerical minimization is performed in the cell shown in Fig. \ref{fig2}.
In order to minimize finite-size effects, the cell height $H$ is taken to be 
at least four times the value of $L$ (note that $\theta_s$ decays exponentially
to $\alpha_\infty$ for $y\gg 2L\cos\alpha/\pi$). We checked that the
value of $\lambda f_I^{ns}$ is independent of $\lambda$, as expected, so the
non-singular contribution to $B(\alpha,w)$, $B_{I,ns}(\alpha)$, is related
to $f_I^{ns}$ via $B_{I,ns}(\alpha)=\lambda f_I^{ns}$. The numerical
results are shown for both the $N^\perp$ and $N^\parallel$ textures 
in Fig. \ref{fig3}. The agreement between the results obtained from the
finite element method and boundary element method is excellent, as for
the evaluation of $\theta_{ns}
(\mathbf{r})$, which takes non-negligible values only above the contour $C_1$
and vanishes close to the wedges and apexes (see Fig. \ref{fig4}).  
Our results show that the non-singular contribution corresponding to the
$N^\perp$ (resp. $N^\parallel$) texture is smaller than the contribution
associated to the $N^\parallel$  (resp. $N^\perp$) texture for $\alpha<\pi/4$ 
(resp. $\alpha>\pi/4$).

\subsection{Exact evaluation of $f_I$}

\begin{figure}
\centerline{\includegraphics[width=.95\columnwidth]{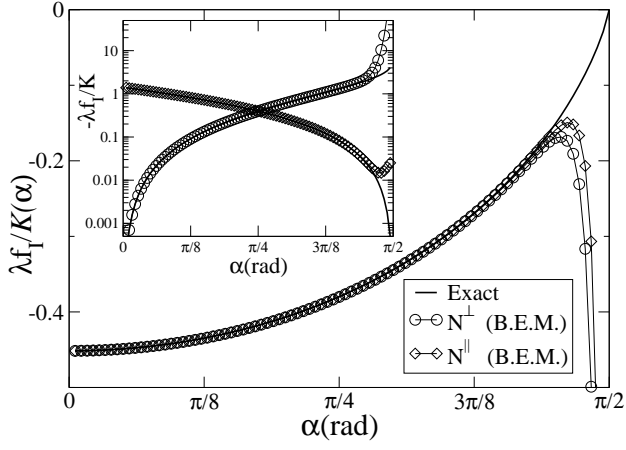}}
\caption
{Plot of $B_{I}/{\cal K}(\alpha)$ as a function of $\alpha$. The 
continuous broad line corresponds to the exact expression, and the symbols
correspond to the estimates from the numerical data for $f_{I,ns}$ (from
the boundary element method): circles for the $N^\perp$ texture and diamonds
for the $N^\parallel$ texture. Inset: plot of $-B_{I}/K$ as a function
of $\alpha$. The meaning of the symbols is the same as in the main panel.
}
\label{fig5}
\end{figure}

In the previous paragraphs we have outlined how to obtain 
$\theta(\mathbf{r})=\theta_{s}(\mathbf{r})+\theta_{ns}(\mathbf{r})$, and
from that, to obtain $f_I=f_{I}^{s}+f_{I}^{ns}$, as well as $B_I(\alpha)= 
B_{I,s}(\alpha)+B_{I,ns}(\alpha)$. However, it is possible to obtain 
$f_I$ directly without knowing the explicit form of $\theta(\mathbf{r})$
by using a Schwarz-Christoffel transformation \cite{barbero1,davidson2}:
\begin{eqnarray}
&&z=\int \dd \bar{\zeta} \frac{C}{(\tilde{\zeta}+1)^{1/2-\alpha/\pi}
\tilde{\zeta}^{2\alpha/\pi}(\tilde{\zeta}-1)^{1/2-\alpha/\pi}}
\nonumber\\
&&
=C'\zeta^{1-
\frac{2\alpha}{\pi}} {_2F_1}\left(\frac{1}{2}-\frac{\alpha}{\pi},\frac{1}{2}
-\frac{\alpha}{\pi},\frac{3}{2}-\frac{\alpha}{\pi},\zeta^2\right)+C''
\label{sct}
\end{eqnarray}
where $z=x+iy$, ${_2F_1}(a,b,c,z)$ is the Gauss hypergeometric function 
and $C$, $C'$ and $C''$ complex constants. 
This conformal transformation maps the 
minimization cell for $H\to \infty$ into the upper half $\zeta$-plane (see 
Fig. \ref{fig2}), transforming the origin into the origin and the edges 
$z=\pm L\cos \alpha+iL\sin\alpha$ into $\zeta=\pm 1$, respectively. 
These conditions fix 
the values of $C'$ and $C''$, so the Schwarz-Christoffel transformation 
finally reads:
\begin{eqnarray}
&&z=\frac{L e^{i\alpha}}{\Gamma\left(\frac{3}{2}-\frac{\alpha}{\pi}\right)
\Gamma\left(\frac{1}{2}+\frac{\alpha}{\pi}\right)}\zeta^{1-
\frac{2\alpha}{\pi}}\nonumber\\&&\times 
{_2F_1}\left(\frac{1}{2}-\frac{\alpha}{\pi},\frac{1}{2}
-\frac{\alpha}{\pi},\frac{3}{2}-\frac{\alpha}{\pi},\zeta^2\right)
\label{sct2}
\end{eqnarray}
where $\Gamma(x)$ is the gamma function. Eq. (\ref{sct2}) can be formally 
inverted, so $\zeta=\zeta(z/L;\alpha)=x'(x/L,y/L;\alpha)+iy'(x/L,y/L;\alpha)$.
We will consider the limit $\eta/L\to 0$, and $R_c/L$ small but finite. 
In this approach, the boundary of zone $II$ becomes under the 
Schwarz-Christoffel transformation the real axis in the $\zeta$-plane, 
rounded around $\zeta=0$ and $\pm 1$. The expansion of Eq. (\ref{sct2}) around
these values show that the circles of radii $R_c$ around the origin
and the edges $z=\pm L\cos \alpha + iL\sin \alpha$ map into circles (up to 
corrections of order of $(R_c/L)^2$) of radii $\epsilon_1$ for $\zeta=\pm 1$
and $\epsilon_2$ for $\zeta=0$ given by:
\begin{eqnarray}
&&\epsilon_1=\frac{1}{2}\left(\frac{1+\frac{2\alpha}{\pi}}
{1-\frac{2\alpha}{\pi}}\right)^{\frac{1}{\frac{1}{2}+\frac{\alpha}{\pi}}}
\left(\frac{R_c}{L}\Gamma\left[\frac{3}{2}-\frac{\alpha}{\pi}
\right]\Gamma\left[\frac{1}{2}+\frac{\alpha}{\pi}\right]\right)^{\frac{1}
{\frac{1}{2}+\frac{\alpha}{\pi}}}\nonumber
\\
&&\epsilon_2=\left(\frac{R_c}{L}\Gamma\left[\frac{3}{2}-\frac{\alpha}{\pi}
\right]\Gamma\left[\frac{1}{2}+\frac{\alpha}{\pi}\right]\right)^{\frac{1}{1-
\frac{2\alpha}{\pi}}}\label{epsilon12}
\end{eqnarray} 
As the Schwarz-Christoffel transformation is conformal, and 
$\theta(\mathbf{r})$ is harmonic, we are going to find the solution to the
Laplace equation in the half $\zeta$-plane,  $\tilde\theta$, subject to the 
boundary conditions $\tilde\theta=\alpha_\infty$ for $|x'|>
1+\epsilon_1$, and $\tilde\theta=\alpha_\infty+(x'/|x'|)(\alpha-\alpha_\infty)$
for $\epsilon_2<|x'|<1-\epsilon_1$. The solution $\tilde\theta(x',y')$ in
the image of the region $I$ on the $\zeta$-plane, $I'$, is
given by \cite{barbero1}:
\begin{eqnarray}
&&\tilde\theta(x',y')=\alpha_\infty+\frac{\alpha-\alpha_\infty}{\pi}\arctan
\frac{y'}{x'-1}\nonumber\\
&&-\frac{2(\alpha-\alpha_\infty)}{\pi}\arctan\frac{y'}{x'}+
\frac{\alpha-\alpha_\infty}{\pi}\arctan
\frac{y'}{x'+1}
\label{tildetheta}
\end{eqnarray}
From the solution $\tilde\theta$, we can
obtain $\theta(x,y)=\tilde\theta[x'(x/L,y/L;\alpha),y'(x/L,y/L;\alpha)]$.  
We note that we do not have an explicit expression for $x'$ and $y'$
as functions of $x$ and $y$, so we cannot give an analytic expression for
$\theta(x,y)$. However, we can evaluate exactly $f_I$ since:
\begin{eqnarray}
&&f_I=\frac{K}{2\lambda}\int_{I} \dd x \dd y 
|\boldsymbol \nabla \theta|^2=\frac{K}{2\lambda}\int_{I'} \dd x' \dd y'
|\boldsymbol \nabla' \tilde \theta|^2 
\nonumber\\
&&=\frac{K}{2\lambda} 
\int_{\cal B} \tilde \theta (\boldsymbol \nu' \cdot \boldsymbol
\nabla' \tilde \theta) \dd \mathbf{s}'
\label{mfofreeenergy2}
\end{eqnarray}
where $\cal B$ is the image in the $\zeta$-plane of the boundary of the zone 
$I$. After an straightforward calculation, we get the expressions for $f_I$
and $B_I(\alpha)$:
\begin{eqnarray}
&&f_I=\frac{q\mathcal{K}(\alpha)}{2\pi}\Bigg[-\ln \frac{qR_c\cos\alpha}{\pi}
\nonumber\\&&-
\ln\left(\Gamma\left[\frac{3}{2}-\frac{\alpha}{\pi}
\right]\Gamma\left[\frac{1}{2}+\frac{\alpha}{\pi}\right]\right)
\nonumber\\&&-
\left(\frac{1}{2}-\frac{\alpha}{\pi}\right)\ln\left(\frac{\frac{\pi}{2}+\alpha}
{\frac{\pi}{2}-\alpha}\right)\Bigg]
\label{theoreticalfI}\\
&&B_I(\alpha)=-\mathcal{K}(\alpha)\Bigg[\left(\frac{1}{2}-\frac{\alpha}{\pi}\right)\ln\left(\frac{\frac{\pi}{2}+\alpha}
{\frac{\pi}{2}-\alpha}\right)\nonumber\\
&&+
\ln\left(\Gamma\left[\frac{3}{2}-\frac{\alpha}{\pi}
\right]\Gamma\left[\frac{1}{2}+\frac{\alpha}{\pi}\right]\right)\Bigg]
\label{theoreticalfI2}
\end{eqnarray}
Fig. \ref{fig5} shows an excellent agreement between the theoretical 
prediction for $B_I(\alpha)$ from Eq. (\ref{theoreticalfI2}) and the 
results obtained in the previous subsections, except close to $\alpha=\pi/2$.
The latter may be due to numerical uncertainties in the evaluation of
$B_{I,ns}$, since either it diverges ($N^\perp$ texture) or vanishes 
($N^\parallel$ texture) in that limit. As happened for $B_{I,s}$, the
dependence on the texture comes from ${\cal K}(\alpha)$, leading to the
same conclusions about the relative stability of the nematic textures with
the angle $\alpha$, in agreement with the results obtained in Ref. 
\cite{barbero1}. 

\section{Evaluation of $f_{II}$}

In Ref. \cite{romero} it was observed that there is a dependence on $w$ of the
next-to-leading contribution to the surface free-energy density. However, 
in the previous Section we have shown that the contribution from region $I$ 
only depends on $\alpha$. So, we anticipate that this dependence comes from
region $II$, where inhomogoneities of both the nematic order parameter $S$ and
orientational $\theta$ fields are observed.  

\begin{figure}[t]
\centerline{\includegraphics[width=.95\columnwidth]{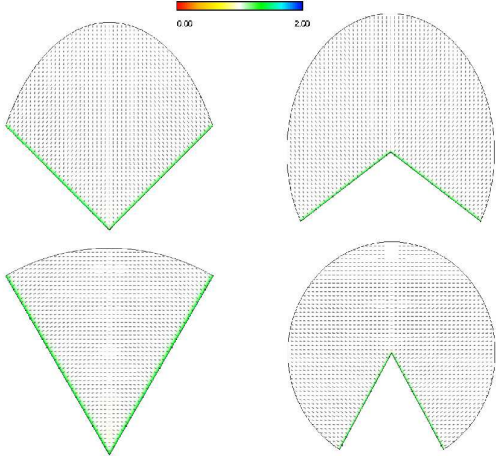}}
\caption
{(Color online)
Typical textures obtained from the full minimization of the LdG function in
region $II$ for $\tau=1$, $\kappa=2$, $w=1.0$, $L=90$ and
$\alpha=\pi/6$ (upper panels) and $\alpha=\pi/3$ (lower panels).
The color map corresponds to the reduced nematic order parameter field
$\tilde S=S/S_b$, and the segments correspond to the nematic director field
$\mathbf{n}$.\label{fig6}}
\end{figure}

The free energy of region $II$ can be evaluated as the sum of the 
contributions of the regions around each cusp (either wedge or apex). In 
each of these regions and if
$R_c\ll L$, we anticipate that the orientational field far from the 
cusp behaves asymptotically as that of the isolated disclination line 
which has been nucleated at the substrate wedge or apex: $\theta \sim I 
\phi$, where $I$ is the topological charge of the disclination line and 
$(r,\phi)$ are
the polar coordinates taking as origin the cusp. The values of $I$ are fixed
by the (strong) anchoring conditions on the substrate and depend on the 
nematic texture: $I_1=-\alpha/(\pi/2-\alpha)$ (resp. $I_1=+1$) 
at the wedge bottom
and $I_2=\alpha/(\pi/2+\alpha)$ (resp. $-(\pi/2-\alpha)/(\pi/2+\alpha)$) at
the apexes for the $N^\perp$ (resp. $N^\parallel$) \cite{barbero1,romero}. 
However, $S$
decreases from the bulk value far from the origin and vanishes as $r\to 0$ in
order to remove the free-energy singularity associated to the defect core.
Although we cannot solve analytically this problem, we can estimate using
an ansatz the free energy associated to this combined distortion of $\theta$ 
and $S$, as shown in the Appendix. However, we resort here to a full 
minimization of the LdG model Eq. (\ref{free_energy}) restricted to region 
$II$. At $r=R_c$, we impose Dirichlet boundary conditions to $\Qvec$, 
where $S$ takes the value corresponding to the order-parameter profile 
corresponding to a planar wall at the distance between the boundary point 
and the closest substrate, the biaxiality parameter $B=0$ and 
$\theta=\alpha_\infty+I(\phi-\pi/2)$. Alternatively we used free boundary
conditions $\boldsymbol \nu \cdot \boldsymbol \nabla \Qvec=0$, leading to
similar results. Fig. \ref{fig6} shows typical textures obtained after 
minimization. At distances $r$ larger than a few correlation lengths, 
$S$ decays to the bulk value except in the neighbourhood of the substrates, 
where takes approximately the value corresponding to the flat substrate 
profile. On the other hand, the orientational field deforms continuously 
in order to satisfy anchoring conditions on each side of the wedge or the apex. 
So, we anticipate that the contribution to the surface free-energy density
$f_{II}$ will scale with $R_c$ as:   
\begin{equation}
\lambda f_{II}=4\sigma_{nw} R_c + {\cal K}(\alpha) \ln R_c + B_{II} 
\label{analisis1}
\end{equation}
The first contribution arises from the inhomogeneities of $S$ close to the
substrates, and the second one from the asymptotic behaviour of $\theta$ for 
$r>1$: by using the Frank-Oseen functional Eq. (\ref{mfofreeenergy}), and
taking into account that $\boldsymbol \nabla \theta=I\mathbf{u}_\phi/r$ (where
$\mathbf{u}_\phi$ is the azimuthal unit vector), we
find that the free-energy contribution is proportional to $(K/2) I^2\Delta 
\phi \ln R_c$, where $\Delta\phi$ is the opening angle of the wedge or apex. 
From this result we get the second contribution in Eq. (\ref{analisis1}) by
noting that ${\cal K}(\alpha)=K I_1^2(\pi-2\alpha)/2+ KI_2^2(\pi+2\alpha)/2$. 
The remaining contribution will give the core free energy per unit length
associated to the disclination lines.

\begin{figure}
\centerline{\includegraphics[width=0.95\columnwidth]{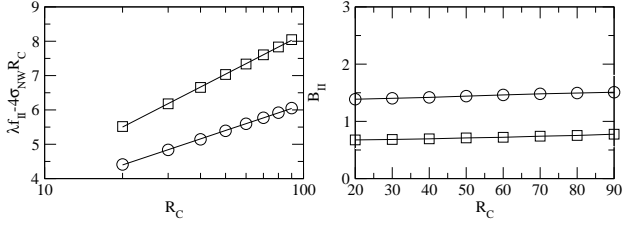}}
\caption
{Left panel: representation of $\lambda f_{II}-4\sigma_{NW} R_c$ as a function
of $R_c$ for $w=1$ and: $\alpha=\pi/6$ (squares) and $\alpha=\pi/3$ (circles).
The straight lines correspond to the logarithmic regressions of the numerical 
data. Right panel: representation of $B_{II}(\alpha,w)$ as a function
of $R_c$. The meaning of the symbols is the same as in the left panel.}
\label{fig7}
\end{figure}

Expression (\ref{analisis1}) can be used to extract $B_{II}$ from
the numerical minimization. So, fixing the value of $\alpha$, and 
for each $w$ (between 0 and 1.5) 
we considered a range of values of $R_c$ 
between $10$ and $90$. The minimization was performed by the mesh-adaptive 
finite-element method used previously \cite{patricio3}, taking $\tau=1$ and
$\kappa=2$ (this choice is motivated to compare with results reported in
the literature \cite{patricio,romero}).
After substrating $4\sigma_{nw}R_c$ to $\lambda f_{II}$, the numerical 
results clearly show a logarithmic dependence on $R_c$, with a slope 
approximately equal to ${\cal K}(\alpha)$ (see left panel in Fig. \ref{fig7}).
Next step was to substract the subdominant contribution 
${\cal K}(\alpha)\ln R_{\infty}$. Now the numerical results are nearly
independent of $R_c$ (see right panel in Fig. \ref{fig7}). The value of
$B_{II}$ is estimated as the mean value of these results, with an errorbar
given by the dispersion of the numerical data around the average.

\section{Discussion}

\begin{figure}
\centerline{\includegraphics[width=0.95\columnwidth]{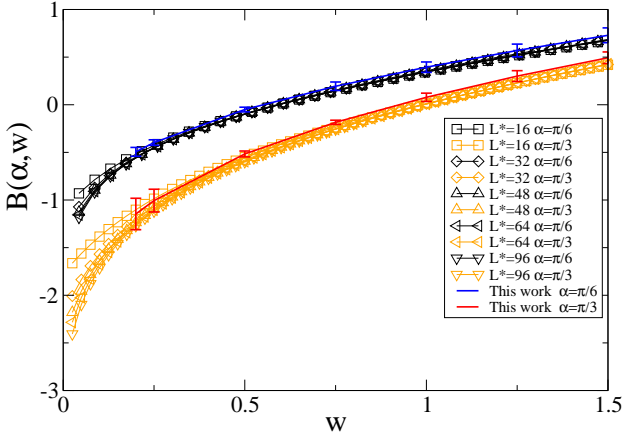}}
\caption{(Color online) 
Comparison between the values of $B(\alpha,w)$ 
obtained from the full minimization of the LdG model \cite{romero} and the 
results obtained in the present work. Symbols correspond to $B(\alpha,w)$ 
obtained from the full minimization of the LdG model for different values
of $L$ and $\alpha=\pi/6$ in the $N^\perp$ texture (black symbols), and
$\alpha=\pi/3$ in the $N^\parallel$ texture (orange or light gray symbols). 
The wide blue (black) and wide red (dark gray) lines with errorbars
correspond to the predictions from this work for $\alpha=\pi/6$ and $\pi/3$,
respectively.}
\label{fig8}
\end{figure}

The results obtained in the previous Sections can be combined as follows:
\begin{equation}
f=f_I+f_{II}+f_{III}=\frac{\sigma_{NW}}{\cos\alpha}+f_{elastic}
\label{finalf0}
\end{equation}
where
\begin{eqnarray}
&&f_{elastic}=\frac{q}{2\pi} {\cal K}(\alpha)\Bigg[-\ln 
\frac{q\cos\alpha}{\pi}\nonumber\\
&&-\left(\frac{1}{2}-\frac{\alpha}{\pi}\right)\ln\left(\frac{\frac{\pi}{2}
+\alpha}{\frac{\pi}{2}-\alpha}\right)\nonumber\\
&&-
\ln\left(\Gamma\left[\frac{3}{2}-\frac{\alpha}{\pi}
\right]\Gamma\left[\frac{1}{2}+\frac{\alpha}{\pi}\right]\right)\Bigg]
\nonumber\\
&&+\frac{q}{2\pi} B_{II}(\alpha,w)
\label{finalf}
\end{eqnarray}
up to corrections of order of $q^2$. Consequently:
\begin{eqnarray}
&&B(\alpha,w)=B_I(\alpha)+B_{II}(\alpha,w)=\nonumber\\
&&
-{\cal K}(\alpha)\Bigg[
\left(\frac{1}{2}-\frac{\alpha}{\pi}\right)\ln\left(\frac{\frac{\pi}{2}
+\alpha}{\frac{\pi}{2}-\alpha}\right)\nonumber\\
&&+
\ln\left(\Gamma\left[\frac{3}{2}-\frac{\alpha}{\pi}
\right]\Gamma\left[\frac{1}{2}+\frac{\alpha}{\pi}\right]\right)\Bigg]
\nonumber\\
&&+ B_{II}(\alpha,w)
\label{finalb}
\end{eqnarray}
This is the main result of our paper. Now we can check this prediction by
comparing these results with those obtained from the full minimization of
the LdG model of a nematic in contact with a sawtoothed substrate 
\cite{romero}. Fig. \ref{fig8} shows the comparison between the results 
reported in the Ref. \cite{romero} ($\tau=1$, $\kappa=2$) 
and the calculated ones in this paper for $\alpha=\pi/6$ ($N^\perp$ texture) 
and $\alpha=\pi/3$ ($N^\parallel$ texture). For large $w$, the agreement
is good, although our results slightly overestimate those from the
full minimization. On the other hand, for small $w$ the curves obtained 
from the full minimisation
converge towards our prediction as $L$ increases. So, our approximation is
accurate even for moderate values of $L$, despite the assumptions 
involved in our approach. Consequently, the scheme considered in this paper
is an alternative to the full minimization technique, which is quite expensive 
from a computational point of view, when $wL>1$.

Finally, it is interesting to note that the elastic contribution to the 
surface free-energy density, Eq. (\ref{finalf}), can be expressed as follows:
\begin{equation}
\lambda f_{elastic}=2\times \left(2\pi K \frac{I_1}{2} \frac{I_2}{2}
\ln\frac{\gamma(\alpha)}{L}\right)+ B_{II}(w,\alpha)
\label{felastic2}
\end{equation}
where we note that ${\cal K}(\alpha)=-4 \pi K (I_1/2) (I_2/2)$, with 
$I_1$ and $I_2$ being the
topological charges associated to the disclination lines at the wedges and
apexes, respectively. The characteristic length $\gamma(\alpha)$,
which absorbs the contribution to $B$ from region $I$, is defined as: 
\begin{equation}
\gamma(\alpha)=\Gamma\left[\frac{3}{2}-\frac{\alpha}{\pi}
\right]\Gamma\left[\frac{1}{2}+\frac{\alpha}{\pi}\right]
\left(\frac{\frac{\pi}{2}
+\alpha}{\frac{\pi}{2}-\alpha}\right)^{\frac{1}{2}-\frac{\alpha}{\pi}}
\label{defgamma}
\end{equation}
This lengthscale decays continuously from $\pi/2$ for $\alpha\to 0$ to $1$
for $\alpha\to \pi/2$. 
We can understand the first contribution in 
Eq. (\ref{felastic2}) as the interaction between
the disclination line at $x=0$ with half its topological charge and the apexes 
at $x=\pm L\cos\alpha$, again with half their topological charges 
\cite{chaikin}. 
As the absolute values of the topological charges are always smaller or equal
than $1$, the nucleation of disclinations at the cusps of the surface is 
favourable with respect to bulk disclinations (which only may have 
half-integer values). With this intepretation, $\gamma(\alpha)$ can be used 
to define an effective core radius $\gamma(\alpha)$, 
and the sum of the core energies is given by $B_{II}(\alpha,w)$.  

\section{Conclusions}

In this paper we have analyzed elastic contribution to the surface free-energy
density for a nematic in contact to sawtoothed substrates in the strong 
anchoring regime, i.e. $wL>1$. We have extended the analysis done in Ref. 
\cite{romero}. So, in addition to the leading 
contribution proportional to $-q\ln q$, with $q$ being the substrate 
periodicity wavenumber $q$, we have characterized the next-to-leading 
term. This term has two contributions: one associated to the deviation of the 
orientational field with respect to the contribution of the array 
of disclination lines nucleated at the cusps of the substrate, and the
core free-energy associated to them. We anticipate that
our analysis can be generalized for other substrate shapes 
when the nematic texture presents topological defects induced by the 
structure, as, for example, 
in crenellated substrates \cite{nuno2}. Furthermore, 
our results may be used to predict accurately
the location of first-order wetting transitions in nematic liquid crystals  
in contact to general substrates \cite{patricio,patricio2,nuno2}. However,
we must note that our analysis is restricted to bulk nematics liquid crystals 
with in-plane deformations in presence of grooved substrates. The effect of 
the substrate structure in the nematic texture in partially filled 
configurations (i.e. with an isotropic fluid in bulk), the influence
of twist nematic deformations \cite{patricio5} or the effect of the structure 
in the $z$ direction deserve further study, but this is beyond the scope of 
this work. 

\acknowledgments

The authors wish to thank Dr. P. Patr\'{\i}cio and Dr. N. M. Silvestre for
very stimulating conversations and technical advice on the finite-element 
method, and Prof. M. M. Telo da Gama, Prof. L. F. Rull and Prof. A. O. Parry 
for enlightening discussions. We acknowledge the 
support from MICINN (Spain) through Grant
FIS2009-09326, and Junta de Andaluc\'{\i}a (Spain) through Grant No. 
P09-FQM-4938, both co-funded by EU FEDER. 

\appendix
\section{\emph{Ansatz} for the evaluation of $f_{II}$}

In order to estimate the free energy associated to the region $II$, we can
make use of an ansatz. As argued in the paper, the nematic order parameter $S$ 
must vanish at the cusps of the substrate to avoid the divergence of the free
energy, but it must converge to the bulk value far from the wedges and apexes.
So, in the region around each cusp, we can suppose that $S$ depends only on 
the radial distance $r$ from the wedge or apex. In particular, we use the 
following ansatz \cite{tasinkevych}: $S(r)=S_b(1-e^{-\frac{r}{\beta}})$,
where $S_b$ is the bulk nematic order parameter and $\beta$ is a lengthscale 
to be determined later. We will assume that the biaxiality parameter $B=0$ 
everywhere. Regarding the orientational field $\theta$, we assume
that has the expression corresponding to a disclination line placed at the
origin: $\theta=\alpha_\infty+I(\phi-\pi/2)$, where $I$ is the winding number
associated to the disclination line, and $\alpha_\infty$ is the far-field value
of the orientational field ($0$ for the $N^\perp$  texture, and $\pi/2$ for
the $N^\parallel$ texture). 
Substituting this \emph{ansatz} in the LdG functional Eq. (\ref{free_energy}),
this expression reduces to a function of $\beta$:
\begin{eqnarray}
&&\lambda f\approx\frac{13}{144}(\pi\mp2\alpha)\beta^{2}-2w(R_c-\beta)
+\frac{1}{7}\Bigg\{9I^{2}(\pi\mp2\alpha)\Big[\gamma
\nonumber\\&&+\ln\Big(\frac{R_c}{2\beta}\Big)\Big]+
\frac{11}{16}(\pi\mp2\alpha)
+\frac{1}{8}\Big(\frac{\sin[(I-1)(\pi\mp2\alpha)]}{(I-1)}\nonumber\\&&
\mp 18I\sin(2\alpha)
-3\big[\sin(-2I(\pi\mp\alpha)+(I-1)\pi)\nonumber\\
&&-\sin(-2I(\pm\alpha)+(I-1)\pi)\big]\Big)\Bigg\}
\label{free_energy_ansatz}
\end{eqnarray}
where we have considered the case $\tau=1$ and $\kappa=2$ (for other
values, we can get straightforwardly analogous expressions). In this 
expression, the upper sign corresponds to the wedge situation (so $I=I_1$),
and the lower one to the apex (so $I=I_2$), and $\gamma$ is the Euler 
constant. Minimizing Eq. (\ref{free_energy_ansatz}) with respect to $\beta$,
we obtain the optimal value for this lengthscale:
\begin{equation}
\beta=-\frac{72}{13}\frac{w}{(\pi\mp2\alpha)}+\sqrt{\left(\frac{72}{13}
\frac{w}{(\pi\mp2\alpha)}\right)^{2}+\frac{648}{91}I^{2}}
\label{ansatz17}\end{equation}
Substituting Eq. (\ref{ansatz17}) into Eq. (\ref{free_energy_ansatz}), we
get the free energy estimate associated to each region around a cusp, and
provides an upper limit to the real value of $\lambda f$. 

\begin{figure}
\centerline{\includegraphics[width=0.95\columnwidth]{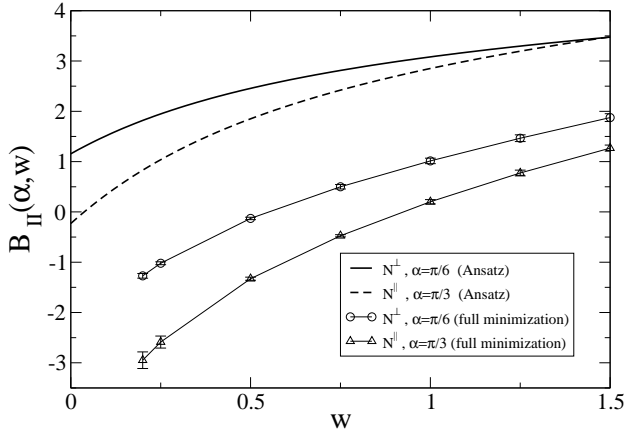}}
\caption{ 
Comparison between the values of $B_{II}(\alpha,w)$ 
obtained from the full minimization of the LdG model and the 
\emph{ansatz} (see text). Symbols correspond to $B_{II}(\alpha,w)$ 
obtained from the full minimization of the LdG model 
for $\alpha=\pi/6$ in the $N^\perp$ texture (circles), and
$\alpha=\pi/3$ in the $N^\parallel$ texture (up triangles).  
The wide continuous and dashed lines correspond to the estimates from
the \emph{ansatz} for $\alpha=\pi/6$ and $\alpha=\pi/3$, respectively.}
\label{fig9}
\end{figure}

In order to have an estimate of $B_{II}$, we substract to the obtained
values of $\lambda f$ through this ansatz the surface and elastic
contributions $-2w R_c$ and $(K/2) I^2 (\pi\mp 2\alpha)\ln R_c=(9/7) I^2 
(\pi\mp 2\alpha)\ln R_c$, respectively. After this, the estimate of $B_{II}$ is
obtained as the sum of the results obtained for the wedge and apex.
Fig. \ref{fig9} shows the comparison of this estimates with the results
obtained from the full minimization, for  $\alpha=\pi/6$ ($N^\perp$ texture)
and $\alpha=\pi/3$ ($N^\parallel$) texture. They have the same order of 
magnitude, although our \emph{ansatz} overestimate the value of $B_{II}$, 
and qualitatively the dependence of $B_{II}$ with $w$ is recovered.


\begin{thebibliography}{33}
\expandafter\ifx\csname natexlab\endcsname\relax\def\natexlab#1{#1}\fi
\expandafter\ifx\csname bibnamefont\endcsname\relax
  \def\bibnamefont#1{#1}\fi
\expandafter\ifx\csname bibfnamefont\endcsname\relax
  \def\bibfnamefont#1{#1}\fi
\expandafter\ifx\csname citenamefont\endcsname\relax
  \def\citenamefont#1{#1}\fi
\expandafter\ifx\csname url\endcsname\relax
  \def\url#1{\texttt{#1}}\fi
\expandafter\ifx\csname urlprefix\endcsname\relax\def\urlprefix{URL }\fi
\providecommand{\bibinfo}[2]{#2}
\providecommand{\eprint}[2][]{\url{#2}}

\bibitem{lee} B.-W. Lee and N. A. Clark, Science {\bf 291}, 2576 (2001).
\bibitem{kim} J.-H. Kim, M. Yoneya and H. Yokoyama, Nature {\bf 420}, 19 
(2002).
\bibitem{ferjani} S. Ferjani, Y. Choi, J. Pendery, R. G. Petschek and C. 
Rosenblatt, Phys. Rev. Lett. {\bf 104}, 257801 (2010).
\bibitem[{\citenamefont{Brown et~al.}(2000)\citenamefont{Brown, Towler, Hui,
  and Bryan-Brown}}]{Brown_2000}
\bibinfo{author}{\bibfnamefont{C.~V.} \bibnamefont{Brown}},
  \bibinfo{author}{\bibfnamefont{M.~J.} \bibnamefont{Towler}},
  \bibinfo{author}{\bibfnamefont{V.~C.} \bibnamefont{Hui}}, \bibnamefont{and}
  \bibinfo{author}{\bibfnamefont{G.~P.} \bibnamefont{Bryan-Brown}},
  \bibinfo{journal}{Liq. Cryst.} \textbf{\bibinfo{volume}{27}},
  \bibinfo{pages}{233} (\bibinfo{year}{2000}).

\bibitem[{\citenamefont{Uche et~al.}(2005)\citenamefont{Uche, Elston, and
  Parry-Jones}}]{Parry-Jones1}
\bibinfo{author}{\bibfnamefont{C.}~\bibnamefont{Uche}},
  \bibinfo{author}{\bibfnamefont{S.~J.} \bibnamefont{Elston}},
  \bibnamefont{and} \bibinfo{author}{\bibfnamefont{L.~A.}
  \bibnamefont{Parry-Jones}}, \bibinfo{journal}{J. Phys. D: Appl. Phys.}
  \textbf{\bibinfo{volume}{38}}, \bibinfo{pages}{2283} (\bibinfo{year}{2005}).

\bibitem[{\citenamefont{Uche et~al.}(2006)\citenamefont{Uche, Elston, and
  Parry-Jones}}]{Parry-Jones2}
\bibinfo{author}{\bibfnamefont{C.}~\bibnamefont{Uche}},
  \bibinfo{author}{\bibfnamefont{S.~J.} \bibnamefont{Elston}},
  \bibnamefont{and} \bibinfo{author}{\bibfnamefont{L.~A.}
  \bibnamefont{Parry-Jones}}, \bibinfo{journal}{Liq. Cryst.}
  \textbf{\bibinfo{volume}{33}}, \bibinfo{pages}{697} (\bibinfo{year}{2006}).

\bibitem{davidson} A. J. Davidson, C. V. Brown, N. J. Mottram, S. Ladak and
C. R. Evans, Phys. Rev. E {\bf 81}, 051712 (2010).

\bibitem{evans} C. R. Evans, A. J. Davidson, C. V. Brown and N. J. Mottram,
J. Phys. D: Appl. Phys. {\bf 43}, 495105 (2010).

\bibitem{nuno} N. M. Silvestre, P. Patr\'{\i}cio and M. M. Telo da Gama,
Phys. Rev. E {\bf 69}, 061402 (2004). 

\bibitem{ohzono} T. Ohzono and J.-i. Fukuda, Nat. Commun. {\bf 3}, 701 (2012).

\bibitem{berreman}D. W. Berreman, Phys. Rew. Lett. {\bf 28}, 1683 (1972)

\bibitem{gennes}P.G. de Gennes and J. Prost, \emph{The Physics of Liquid
Crystals}, 2nd ed. (Oxford University Press, Oxford, 1995)

\bibitem{barbero1}G. Barbero, Lett. Nuovo Cimento Soc. Ital. Fis. {\bf 29},
553 (1980)

\bibitem{barbero2}G. Barbero, Lett. Nuovo Cimento Soc. Ital. Fis. {\bf 32},
60 (1981)

\bibitem{barbero3}G. Barbero, Lett. Nuovo Cimento Soc. Ital. Fis. {\bf 34},
173 (1982).

\bibitem[{\citenamefont{Kitson and Geisow}(2002)}]{Kitson_2002}
\bibinfo{author}{\bibfnamefont{S.}~\bibnamefont{Kitson}} \bibnamefont{and}
  \bibinfo{author}{\bibfnamefont{A.}~\bibnamefont{Geisow}},
  \bibinfo{journal}{Appl. Phys. Lett.} \textbf{\bibinfo{volume}{80}},
  \bibinfo{pages}{3635} (\bibinfo{year}{2002}).

\bibitem[{\citenamefont{Fukuda et~al.}(2007)\citenamefont{Fukuda, Yoneya, and
  Yokoyama}}]{Fukuda}
\bibinfo{author}{\bibfnamefont{J.~I.} \bibnamefont{Fukuda}},
  \bibinfo{author}{\bibfnamefont{M.}~\bibnamefont{Yoneya}}, \bibnamefont{and}
  \bibinfo{author}{\bibfnamefont{H.}~\bibnamefont{Yokoyama}},
  \bibinfo{journal}{Phys. Rev. Lett.} \textbf{\bibinfo{volume}{98}},
  \bibinfo{pages}{187803} (\bibinfo{year}{2007}).

\bibitem{patricio5} P. Patr\'{\i}cio, M. M. Telo da Gama, and S. Dietrich, 
Phys. Rev. Lett. 88, 245502 (2002)

\bibitem{harnau} L. Harnau, S. Kondrat and A. Poniewierski, Phys. Rev.
E {\bf 72}, 011701 (2005)

\bibitem{kondrat} S. Kondrat, A. Poniewierski and L. Harnau, Liq. Cryst. {\bf
32}, 95 (2005).

\bibitem{harnau2} L. Harnau and S. Dietrich, Europhys. Lett. {\bf 73}, 28 
(2006).

\bibitem{harnau3} L. Harnau, S. Kondrat and A. Poniewierski, Phys. Rev. E
{\bf 76}, 051701 (2007). 

\bibitem{barbero4} G. Barbero, A.S. Gliozzi, M. Scalerandi and L.R. Evangelista,
Phys. Rew. E {\bf 77}, 051703 (2008)

\bibitem[{\citenamefont{Yi et~al.}(2009)\citenamefont{Yi, Lombardo, Ashby,
  Barberi, Maclennan, and Clark}}]{Yi_2009}
\bibinfo{author}{\bibfnamefont{Y.}~\bibnamefont{Yi}},
  \bibinfo{author}{\bibfnamefont{G.}~\bibnamefont{Lombardo}},
  \bibinfo{author}{\bibfnamefont{N.}~\bibnamefont{Ashby}},
  \bibinfo{author}{\bibfnamefont{R.}~\bibnamefont{Barberi}},
  \bibinfo{author}{\bibfnamefont{J.~E.} \bibnamefont{Maclennan}},
  \bibnamefont{and} \bibinfo{author}{\bibfnamefont{N.~A.} \bibnamefont{Clark}},
  \bibinfo{journal}{Phys. Rev. E} \textbf{\bibinfo{volume}{79}},
  \bibinfo{pages}{041701} (\bibinfo{year}{2009}).

\bibitem{poniewierski} A. Poniewierski, Eur. Phys. J. E {\bf 31}, 169 (2010).

\bibitem{romero}J. M. Romero-Enrique, C.-T. Pham and P.
Patr\'{\i}cio, Phys. Rev. E {\bf 82}, 011707 (2010).

\bibitem{bramble} J. P. Bramble, S. D. Evans, J. R. Henderson, C. Anquetil,
D. J. Cleaver and N. J. Smith, Liq. Cryst. {\bf 34}, 1059 (2007).

\bibitem{patricio} P. Patr\'{\i}cio, C.-T. Pham and J. M. Romero-Enrique,
Eur. Phys. J. E {\bf 26}, 97 (2008).

\bibitem{patricio2} P. Patr\'{\i}cio, J. M. Romero-Enrique, 
N. M. Silvestre, N. R. Bernardino and M. M. Telo da Gama,
Mol. Phys. {\bf 109}, 1067 (2011).

\bibitem{patricio4} P. Patr\'{\i}cio, N. M. Silvestre, C.-T. Pham and J. M. 
Romero-Enrique, Phys. Rev. E {\bf 84}, 021701 (2011).

\bibitem{zienkiewicz}O.C. Zienkiewicz and R.L. Taylor, \emph{The Finite Element
Method}, 5th ed. (Butterworth-Heineman, Oxford, 2000).

\bibitem{patricio3} P. Patr\'{\i}cio, M. Tasinkevych and M.M.
Telo da Gama, Eur. Phys. J. E {\bf 7}, 117 (2002).

\bibitem{brebbia} C. A. Brebbia and J. Dom\'{\i}nguez, \emph{Boundary elements:
an introductory course} 2nd ed. (Computational Mechanics Publications, 
Southampton, 1992). 

\bibitem{katsikadelis}J.T. Katsikadelis, \emph{Boundary Elements: Theory and
Applications} (Elsevier, Amsterdam, 2002)

\bibitem{tasinkevych}M. Tasinkevych, N.M. Silvestre, P. Patricio and M.M.
Telo da Gama, Eur. Phys. J. E {\bf 9}, 341 (2002)

\bibitem{chaikin}P.M. Chaikin y T.C. Lubensky, \emph{Principles of Condensed
Matter Physics}, (Cambridge University Press, Cambridge, 1997)

\bibitem{davidson2} A. J. Davidson and N. J. Mottram, Eur. J. Appl. Math. 
{\bf 23}, 99 (2011).

\bibitem{nuno2} N. M. Silvestre, Z. Eskandari, P. Patr\'{\i}cio, 
J. M. Romero-Enrique and M. M. Telo da Gama, Phys. Rev. E (submitted) (2012).

\end{thebibliography}
%

\newpage

\end{document}